\documentclass[12pt]{article}

\usepackage{cite}
\usepackage{graphicx}
\usepackage{epsfig}
\usepackage{amsfonts}
\usepackage{amssymb}
\usepackage{amsmath,amssymb}

\usepackage{epsf,epsfig}

\date{\today}

\newcommand{\si}{\sigma}

\newcommand{\ee}{\end{equation}}
\newcommand{\eea}{\end{eqnarray}}
\newcommand{\be}{\begin{equation}}
\newcommand{\bea}{\begin{eqnarray}}

\begin{document}
\begin{center}

{\Large \bf  Reissner-Nordstr\"om black holes
with non-Abelian hair}
\vspace{0.6cm}
\\
Carlos Herdeiro$^{\dagger}$, 
Vanush Paturyan$^{\ddagger}$,
 Eugen Radu$^{\dagger}$
and  D. H. Tchrakian$^{\ddagger \star}$
\vspace{0.3cm}
\\
$^{\dagger}${\small  Departamento de F\'isica da Universidade de Aveiro and CIDMA, 
\\
 Campus de Santiago, 3810-183 Aveiro, Portugal}
\\
$^{\ddagger}${\small Department of Computer Science,
National University of Ireland Maynooth}
\\
$^{\star}${\small School of Theoretical Physics -- DIAS, 10 Burlington
Road, Dublin 4, Ireland }
\end{center}
\begin{abstract}

We consider $d\geqslant 4$ Einstein--(extended-)Yang-Mills theory, where the gauge sector is augmented by higher order terms. Linearizing the  (extended)
Yang-Mills equations  on the background of the electric Reissner-Nordstr\"om  (RN) 
black hole, we show the existence of normalizable zero modes, dubbed  \textit{non-Abelian  
magnetic stationary clouds}. The non-linear realization of
these clouds bifurcates the RN family into a branch of static, spherically symmetric, electrically charged and asymptotically flat black holes with non-Abelian hair. Generically, the hairy black holes 
are thermodynamically preferred over the RN solution, which, in this model, becomes unstable against the formation of non-Abelian hair, for sufficiently large values of the electric charge.  

\end{abstract}

\newpage

\section{Introduction}
According to the uniqueness theorems~\cite{Chrusciel:2012jk}, the Kerr-Newman solution~\cite{Newman:1965my} is the most general asymptotically flat, non-singular (on and outside the event horizon), single black hole (BH) solution in electro-vacuum General Relativity (GR). Such theorems have shaped the appealing worldview that the multitude of BHs in the Cosmos is well described by the Kerr metric~\cite{Kerr:1963ud}, when near equilibrium (assuming their electric charge is negligible). 
This worldview, however, relies deeply on another ingredient: the celebrated BH perturbation theory, developed in the 1970s. This framework allowed establishing that uncharged rotating [charged non-rotating] BHs, described by the Kerr [Reissner-Nordstr\"om (RN)]  metric, are stable against vacuum~\cite{Whiting:1988vc} [electro-vacuum~\cite{Moncrief:1974gw,Moncrief:1974ng}] linear perturbations, in a mode analysis.\footnote{Extremal BHs (both Kerr and RN) were outside the scope of these proofs and have been recently shown to be unstable against linear perturbations~\cite{Aretakis:2012ei}. The stability of the Kerr-Newman BH is still under study -- see $e.g.$~\cite{Pani:2013ija,Zilhao:2014wqa}.} 

A follow-up question is if Kerr or RN BHs are still stable when \textit{other} matter fields are considered (beyond electro-vacuum). The relevance of this question is illustrated by the \textit{superradiant instability} of Kerr BHs, triggered by massive bosonic fields, which are most commonly taken to be scalar or Proca fields (see~\cite{Brito:2015oca} for a review). In the presence of appropriate seeds of these fields, the instability develops, extracting rotational energy and angular momentum from the BH, that piles up into a bosonic cloud around the horizon. For a single superradiant mode, this growth stops when the BH's angular velocity decreases sufficiently to synchronise with the phase angular velocity of the superradiant mode~\cite{East:2017ovw}, forming a \textit{Kerr BH with synchronised bosonic hair}~\cite{Herdeiro:2014goa,Herdeiro:2015gia,Herdeiro:2015tia,Herdeiro:2016tmi}. The existence of these new  BH solutions, that bifurcate from the Kerr solution, could be inferred prior to this dynamical analysis, by analysing the linearised bosonic wave equation around the Kerr BH and observing the existence of zero modes, dubbed \textit{stationary bosonic clouds}, precisely at the threshold of the unstable modes~\cite{Hod:2012px,Herdeiro:2014goa,Hod:2013zza,Benone:2014ssa,Hod:2014baa,Hod:2016yxg,Hod:2016lgi}. The hairy BHs are the non-linear realisation of these zero modes~\cite{Herdeiro:2014ima} and they are thermodynamically favoured over Kerr BHs~\cite{Herdeiro:2014goa}. 

The above considerations for the Kerr case can be extended to the generic Kerr-Newman case: the superradiant instability exists~\cite{Furuhashi:2004jk}, there are zero-modes~\cite{Hod:2013zza,Benone:2014ssa} and Kerr-Newman BHs with scalar hair have been constructed~\cite{Delgado:2016jxq}. But the same does \textit{not} apply in the particular case of the RN BH. Even though superradiant \textit{scattering} exists around RN BHs, these are not afflicted by the superradiant \textit{instability} of massive bosonic fields.\footnote{This has been proven for scalar fields in~\cite{Hod:2013eea,Hod:2013nn}. Still, an instability can be obtained by confining the bosonic field in a box around the RN BH~\cite{Herdeiro:2013pia,Degollado:2013bha}, with an analogous dynamical development of the instability to that seen in the Kerr case~\cite{Sanchis-Gual:2015lje,Sanchis-Gual:2016tcm}.} As such, massive bosonic hair does not grow around the asymptotically flat RN BH, in contrast to the Kerr case, and in agreement with known no-hair theorems~\cite{Mayo:1996mv}. The purpose of this paper is to point out that in the presence of a class of  non-Abelian fields, it is possible to grow hair around the asymptotically flat RN BH, and the process resembles the aforementioned discussion of the Kerr superradiant instability. 

\bigskip

BHs with non-Abelian hair were initially discovered in $d=4$ Einstein--Yang-Mills  (EYM) $SU(2)$ theory \cite{89}. These so called \textit{coloured} BHs are asymptotically flat and their global YM charge is completely screened, endowing them with a \textit{single} global ``charge" -- the ADM mass. Since for a given value of the mass there can be infinitely many
different solutions, the no-hair conjecture is violated. 
This discovery  
 triggered an extensive search for hairy BHs
in various other models -- see \cite{Volkov:1998cc,Herdeiro:2015waa,Kleihaus:2016rgf,Volkov:2016ehx} 
for reviews.\footnote{The original examples of BHs with non-Abelian hair~\cite{89}
are not known in closed form.
 More recent investigations of
 supersymmetric models with non-Abelian fields
have led
to closed form examples of $d=4$
 asymptotically flat hairy BHs --
see 
\cite{Huebscher:2007hj}
-\cite{Bueno:2014mea}.
}

The coloured BHs in~\cite{89} are, however, unstable against spherical linear perturbations within the EYM model
\cite{Straumann:1990as,Volkov:1995np}.
This instability can be (partly) attributed to the fact that 
 they possess a purely magnetic gauge field and to the absence of a global YM charge ($a.k.a$ 
`baldness' theorem
\cite{Galtsov:1989ip,Ershov:1991nv,Bizon:1992pi}).
This also signifies that the 
(EYM embedded, Abelian) RN BHs and the coloured BHs in \cite{89} 
form disconnected `branches' of the EYM $SU(2)$ model.\footnote{This `baldness' result can be circumvented
by considering a larger gauge group~\cite{Galtsov:1991au}, which allows for $electrically$ charged coloured BHs in EYM theory. These solutions
are, however, superpositions of a RN BH and the $SU(2)$ BHs in \cite{89},
the electric charge being carried by the $U(1)$ subgroup of the larger gauge group.
Consequently, they do not violate the spirit of the `baldness' theorem and are also perturbatively unstable.}

Analogous coloured BHs, with a purely magnetic field, perturbatively unstable
 and with a solitonic limit, exist in more than four spacetime dimensions~\cite{Brihaye:2002hr,Brihaye:2002jg,Radu:2005mj}.  For $d>4$, however, a Derrick-type virial argument implies that 
no finite mass solutions can be found 
in standard EYM theory.
The path taken in~\cite{Brihaye:2002hr,Brihaye:2002jg,Radu:2005mj}, 
was to extend the YM action with particular 
 higher order terms which yield a YM
version of Lovelock's gravity~\cite{Lovelock:1971yv}. The resulting Einstein-\textit{extended}-YM (EeYM) model~\cite{Tchrakian:1984gq} has the desirable property that equations of motion are still second order and Ostrogradsky instabilities~\cite{Ostrogradsky:1850fid} are avoided.

In $d\geqslant 4$ EeYM, it turns out to be possible to circumvent
the no-go results in~\cite{Galtsov:1989ip,Ershov:1991nv,Bizon:1992pi} and obtain coloured, electrically charged BHs \textit{continuously connected to the RN solution} (embedded in this model)~\cite{Radu:2011ip}.
Moreover, as we shall show: 1) the RN solution becomes unstable against eYM perturbations; 2) threshold eYM linear perturbations correspond to static eYM clouds on the RN background (as test fields); and 3) the non-linear realisation of these clouds  corresponds to coloured, electrically charged BH solutions in EeYM theory, which are thermodynamically favoured over the  RN BHs. Thus, we argue, eYM ``matter" triggers an instability of RN BHs that parallels the familiar superradiant instability of Kerr BHs, likely leading to a similar outcome: the dynamical formation of a RN BH with non-Abelian hair, of the type we present below.

\section{The model}
\subsection{The general framework}

The $d\geqslant 4$, EeYM action reads:
\begin{eqnarray}
\label{action}
\mathcal{S}=-\frac{1}{16\pi G}\int_{\mathcal{M}} d^d x \sqrt{-g}
\left(
R
-\sum_{p=1}^P  
\tau_{(p)}~ 
{\mbox {\rm Tr}\ }\{F(2p)^2\}\
\right),
\end{eqnarray}
where $G$,
that will be set to unity, is Newton's constant
and
 $\tau_{(p)}$ are a  set of $P$-input constants
whose values are not constrained \textit{a priori}.

The $2p$-form $F(2p)$ is the $p$-fold antisymmetrised product
$F(2p)=F\wedge F\wedge...\wedge F$ 
of the YM curvature $2$-form $F_{\mu\nu}=\partial_\mu A_{\nu}-\partial_\nu A_{\mu}$.
The  always present 
 $p=1$ term corresponds to the usual YM action,
$F(2)^2=F_{\mu \nu}F^{\mu \nu}$, with $\tau_{(1)}=1/e^2$ ($e$ being the gauge coupling constant). 
For $p=2$ one finds\footnote{Note the analogy with 
the corresponding expression for the Gauss-Bonnet density in Lovelock gravity.}
$
 F(4)^2=6\left[(F_{\mu\nu}F_{\rho\si})^2-4(F_{\mu\rho}F_{\rho\nu})^2+(F_{\mu\nu}^2)^2\right]
$,
which for $d=4$
can be written in the simpler form
$
 F(4)^2=-\frac{1}{2}(\epsilon_{\mu\nu\rho\sigma}F^{\mu\nu}F^{\rho\si})^2
$.
Similar results are found for higher $p$, with increasingly longer
expressions. In $d$ spacetime dimensions, requiring antisymmetry of $F(2p)$
implies that the highest order curvature term $F(2P)$ has $P=[d/2]$
($i.e.$ in four and five dimensions only the first two YM terms contribute, 
the $p=3$ term starts contributing in $d=6$, $etc$.). 
A review of these aspects can be found in
\cite{Tchrakian:2010ar}.

Apart from providing a natural YM counterparts to Lovelock gravity and its mathematical elegance, 
another reason of interest for this EeYM model 
is the occurrence of such $F(2p)^2$ terms in 
 non-Abelian Born-Infeld theory \cite{Tseytlin:1997csa}
or
in the higher loop corrections
to the $d = 10$ heterotic string low energy effective action
\cite{Polchinski:1998rr}.
Here, however, we adopt 
a `phenomenological' viewpoint
and choose the basic action (\ref{action})
primarily for the purpose of identifying the new features induced by such terms.

\medskip
The field equations are obtained by varying the action (\ref{action}) with respect to the field variables
$g_{\mu\nu}$ and $A_{\mu}$
\begin{eqnarray}
\label{field-eqs}
R_{\mu \nu}-\frac{1}{2} Rg_{\mu \nu}= \frac{1}{2} \sum_{p=1}^P \tau_{(p)}~T_{\mu\nu}^{(p)},
\\
\label{YMsph}
D_{\mu}P^{\mu\nu}=0,~~{\rm with}~~
P^{\mu\nu}=\sum_{p=1}^P  
\tau_{(p)}~ \frac{\partial}{\partial F_{\mu\nu}}{\mbox {\rm Tr}\ }\{F(2p)^2\},
\end{eqnarray}
where we define the $p$-stress tensor pertaining to each term in 
the matter Lagrangian as
\be
T_{\mu\nu}^{(p)}=
\mbox{Tr}\{ F(2p)_{\mu\lambda_1\lambda_2...\lambda_{2p-1}}
F(2p)_{\nu}{}^{\lambda_1\lambda_2...\lambda_{2p-1}}
-\frac{1}{4p}g_{\mu\nu}\ F(2p)_{\lambda_1\lambda_2...\lambda_{2p}}
F(2p)^{\lambda_1\lambda_2...\lambda_{2p}}\} .
\label{pstress}
\ee 
 
The solutions reported herein are spherically symmetric, obtained with the metric \textit{Ansatz}
\be
\label{metric}
ds^2=\frac{dr^2}{N(r)}+r^2d\Omega_{d-2}^2-N(r)\sigma^2(r)dt^2,~~
{\rm with}~~N(r)=1-\frac{2m(r)}{r^{d-3}},
\ee
the function $m(r)$ being related to the local mass-energy density up to some $d-$dependent factor.
$r,t$ are the radial and time coordinate, respectively, while $d\Omega_{d-2}^2$
is the line element of a unit sphere.
The choice of gauge group 
compatible with the symmetries of the line element (\ref{metric}) 
is somewhat flexible.  
In this work we choose to employ chiral
representations,
with a $SO(d+1)$ gauge group.
Then a spherically symmetric gauge field Ansatz  is 
\cite{Radu:2012dk,Brihaye:2011nr}:
\begin{eqnarray}
\label{YMansatz}
A= 
\frac{w(r)+1}{r} \Sigma_{ij}\frac{x^i}{r}\,dx^j
+V(r)\Sigma_{d,d+1}\ dt
,~~{\rm with~~} i,j=1, \dots,d-1\,,
\end{eqnarray}
$\Sigma_{ij}$ being the chiral representation matrices of $SO(d-1)$, and $\Sigma_{d,d+1}$ of the $SO(2)$, subalgebras
in $SO(d+1)$, while $x^i$ are the usual Cartesian coordinates,
being related to the spherical coordinates in (\ref{metric})
as in flat space.

\subsection{The equations and known solutions}

Plugging (\ref{metric}) and (\ref{YMsph})  
into the equations of motion (\ref{field-eqs})
results in\footnote{There is also a constraint Einstein equation which
is, however, a differential consequence of (\ref{eq-m})-(\ref{eq-V})
and we shall not display it here.
Also, to simplify the expressions, some $(d,p)$-factors were absorbed in the expression of $\tau_{(p)}$.
}
%
\begin{eqnarray}
\label{eq-m}
&&m'=\frac{1}{2(d-2)}r^{d-2}\sum_{p=1}^{P}\tau_{(p)}  
W^{p-1}
\left[
(2p-1)(d-2p)
\left(
2p N\frac{w'^2}{r^2}+
(d-[2p+1])\,W
\right)
+2p\frac{V'^2}{\sigma^2}
\right],
\\
\label{eq-s}
&&\frac{  \sigma'}{\sigma} =\frac{2}{(d-2) } \frac{w'^2}{r}
\sum_{p=1}^{P} 
p(2p-1)(d-2p)\tau_{(p)} W^{p-1},
\\
\nonumber
&&\sum_{p=1}^{P} 
\tau_{(p)}p
(d-2p)(2p-1)
\Bigg \{
\frac{d}{dr}\Big( r^{d-4}\sigma W^{p-1}N w' \Big)
-
 \frac{1}{2}(p-1)r^{d-2}\sigma W^{p-2}
\bigg(
\frac{Nw'^2}{r^2}+\frac{(d-2p-1)}{2(p-1)}W
\\
\label{eq-w}
&&
{~~~~~~~~~~~~~~~~~~~~~~~~~~~~~~~~~~~~~~~~~~}
-\frac{1}{(d-2p)(2p-1)}\frac{V'^2}{\sigma^2}
\bigg)
        \frac{2w(w^2-1)}{r^4}
\Bigg \}=0,
\\
&&
\label{eq-V}
\sum_{p=1}^{P}  
\tau_{(p)} p
\frac{d}{dr}
\Big(
r^{d-2}W^{p-1}\frac{V'}{\sigma}
\Big)
=0,
\end{eqnarray}
where we use the shorthand notation
\begin{eqnarray}
\label{defW}
W=\left(\frac{w^2-1}{r^2}\right)^2~.
\end{eqnarray}
The 
$d$-dimensional RN BH is a solution of the model for 
a purely electric field,
in which case only the $p=1$ YM term in (\ref{action})
contributes. It has:
\begin{eqnarray}
\label{RN1}
 & w =\pm 1,~\sigma =1, ~
m =m_{(RN)}=M_0-  \frac{4(d-3)}{d-2}\frac{\tau_{(1)} q^2}{r^{d-3}} ,~
~
V=V_{(RN)}=\frac{q}{r_h^{d-3}}-\frac{q}{r^{d-3}},~~{~~} 
\end{eqnarray}
with $M_0$ and $q$ two integration constants fixing the mass and electric charge,
\begin{eqnarray}
\label{RN2}
M=\frac{1}{2}(d-2){\cal V}_{d-2}M_0,~~Q=4(d-3)\tau_{(1)}{\cal V}_{(d-2)} q, 
\end{eqnarray}
where  ${\cal V}_{(d-2)}$ is the area of the unit $(d-2)$-sphere.
These solutions possess an outer event horizon at $r=r_h$,
with $r_h$ the largest root of the equation $N(r)=0$,
a condition which imposes an 
upper bound on the charge parameter,
 $q\leqslant  q_{(max)}=r_h^{d-3}/(8 \tau_{(1)}\frac{d-3}{d-2})^{1/2}$.
Saturating this condition results in an extremal BH.

The electrically uncharged, coloured BHs of  
\cite{89,Brihaye:2002hr,Brihaye:2002jg,Radu:2005mj} are a second set of solutions. They possess nontrivial magnetic field on and outside the horizon,
while the electric  field vanishes $V(r)=0$.
Their mass is finite being the only global charge, since the YM fields leave no imprint at infinity.
These solutions do not trivialize in the limit 
of zero horizon size, becoming gravitating non-Abelian solitons. 

The $d=4$ BHs in 
\cite{Galtsov:1991au}
are yet another set of solutions of the 
above equations, being found for $\tau_{(2)}=0$
and an $SO(3)\times U(1)$
gauge group.
They possess a nonzero magnetic field, and approach
the limit (\ref{RN1})
in the far field.
Similar to Ref.
\cite{89},
the magnetic potential
$w(r)$ possesses at least a node,
with the absence of solutions where it
 becomes infinitesimally small.

We also
note that  the eqs. (\ref{eq-m})-(\ref{eq-V}) are not affected by the transformation:
\begin{eqnarray}
\label{scale}
r\to \lambda r,~~m(r)\to \lambda^{d-3} m(r),~~V\to  V/\lambda,~~
\tau_{(p)}\to \lambda^{4p-2 }\tau_{(p)},
\end{eqnarray}
while $\sigma$ and $w$ remain unchanged.
Thus, in this way one can always take an arbitrary positive value for one of the constants $\tau_{(p)}$.
In this work this symmetry  
 is used to set $\tau_{(1)}= (d-2)/2$.
 Also,
to simplify some relations,
 we shall introduce $\tau\equiv ((d-2)/8)^2\tau_2/\tau_1^3$.

\section{Zero and unstable EeYM modes on the RN BH}

In contrast to the EYM 
model in \cite{Galtsov:1991au},
the presence of a $p>1$ term in the EeYM action 
leads to a
direct interaction between the electric and magnetic fields,
a feature which holds already in the $d=4$ version of the model. 
This, as we shall see,
makes the RN BH \textit{unstable} when considered as a solution of the full model. At the threshold of the unstable modes, a set of zero modes appear, as we now show.

Let us investigate the existence of a perturbative solution  around the RN BH background,
 with $w(r)=\pm 1+\epsilon w_1(r)+\dots$  
($\epsilon$ being a small parameter).
Similar  perturbative expression are written also for $m,\sigma$ and $V$; 
however, to lowest order,  the equation for $w_1(r)$ decouples,
taking the simple form
\begin{eqnarray}
\label{ec-W}
\frac{d}{dr}\Big( r^{d-4}N w_1' \Big)
-2(d-3)r^{d-6}\left[
1-\frac{4 \tau }{ (d-2)(d-3) }V'^2_{(RN)}
\right]
w_1=0,
\end{eqnarray} 
 with $N=1-2m_{(RN)}/r^{d-3}$.
Observe that  only
the $p=1$ and $p=2$ terms enter this equation; other terms only start to contribute at higher order
in perturbation theory.\footnote{More specifically, due to the presence of the 
factor $W^{p-2}$ multiplying the (square of the) electric charge in  (\ref{eq-w}),
 the contribution of a $p>1$ YM term 
in a perturbative expansion
is of order $\mathcal{O}(\epsilon^{2p-3})$.}

The second term in (\ref{ec-W}) can be seen as providing an effective
mass term,
$\mu^2_{(eff)}=1-\frac{4(d-3)\tau }{ d-2 }\frac{q^2 }{r^{2(d-2)}}$ 
for the gauge potential perturbation
$w_1$.
This mass term becomes strictly positive 
for large $r$, $\mu^2_{(eff)} \to 1$
while
it possesses no definite sign near the horizon.
In fact,  $\mu^2_{(eff)}$ becomes negative for large enough values of the electric charge,
and this turns out to be a necessary condition for the existence of $w_1$ solutions with the correct asymptotic behaviour.\footnote{Multiplying by $w_1$
the eq. (\ref{ec-W}), results in the equivalent form
\begin{eqnarray}
\label{ec-W1}
\nonumber
\frac{d}{dr}\Big( r^{d-4}N w_1 w_1' \Big)
= r^{d-4}N w_1'^2
+ 2(d-3)r^{d-6}\mu^2_{(eff)}
w_1^2.
\end{eqnarray}
Normalizable modes have $w_1$ vanishing at infinity. Then, integrating this equation between horizon and infinity, the left hand side vanishes and the first term in the right hand side is strictly positive. Thus, one finds that $\mu^2_{(eff)}$
is necessarily negative for some range of $r$.
}
Requiring $\mu^2_{(eff)}<0$ at the horizon, together with the existence of a horizon (which puts an upper bound on the electric charge), actually  implies the existence of a  maximal value of the electric charge for a given $\tau$, if one wishes normalizable zero mode perturbations to exist:
\begin{eqnarray}
\label{rel-w1}
 \frac{Q^{(max)}}{\tau^{(d-3)/2}}=\frac{1}{8\pi}\frac{2^{d-3}\sqrt{d-3}}{(d-2)^{(d-5)/2}}{\cal V}_{(d-2)}~.
\end{eqnarray} 
For charges smaller than $Q^{(max)}$, we have found that
the equation  (\ref{ec-W}) 
possesses nontrivial solutions,
with $w_1(r)$ starting from some nonzero value at
the horizon and vanishing at infinity.  
In this study, eq. (\ref{ec-W})
implies the existence of the natural control parameter 
${Q}/{\tau^{(d-3)/2}}$,
other quantities being also expressed in units set by $\tau$.
Then, for a given value of 
this parameter between zero and (\ref{rel-w1}),
the solutions exist for a subset of RN backgrounds, specified $e.g.$ by the ratio 
${M}/{\tau^{(d-3)/2}}$,
which result from the numerical analysis.
This set of solutions
 can be indexed by an integer $n$,
corresponding to the node number of $w_1(r)$.
The results displayed in this work 
correspond to the $n=0$ ($i.e$ fundamental) set of solutions.

Following the terminology for scalar fields,
\cite{Hod:2012px,Herdeiro:2014goa,Benone:2014ssa},
 these configurations with infinitesimally small magnetic fields are dubbed  
{\it non-Abelian stationary clouds around RN BHs}.
The corresponding subset of RN BHs span an {\it existence line}
in the parameter space of solutions.\footnote{A
rigorous
existence proof for 
the existence of solutions of the eqs. (\ref{ec-W})
for a number $d=4$ of spacetime dimensions can be found in~\cite{Radu:2016kgu}.
}
This set is shown below, in Figures 3,4 (the blue dotted line); the plotted results are for $d=4,5$ 
but a similar picture has been found for $d=6,\dots,9$ 
and we expect a similar pattern to occur for any $d$.

Even though eq. (\ref{ec-W}) 
 does not appear to be solvable in terms of known functions,
an approximate expression of the solutions
can be found by 
 using the method of matched asymptotic expansions.
For example, for $d=4$,
the solution in the near horizon region [$w_1^{(h)}$] and in the far field [$w_1^{(inf)}$],
as expressed in terms of the compactified coordinate $x=1-r_h/r$,
reads

\begin{eqnarray}
w_1^{(h)}(x)=b+\frac{2b(r_h^4-2Q^2\tau)x}{r_h^2(r_h^2-Q^2)}+\mathcal{O}(x^2),
~~ \qquad 
w_1^{(inf)}(x)=\frac{J}{r_h}(1-x)+\mathcal{O}[(1-x)^2],
\end{eqnarray}
where $b$ and $J$ are free parameters.
These approximate solutions together with their first derivatives are matched at some intermediate point,
which results in the constraint $3r_h^4-Q^2(r_h^2+4 \tau)=0$.
This condition can be expressed
as a
relation between the event horizon area and the
electric charge of the RN BHs  on the existence line:
 \begin{eqnarray}
\label{AHQu} 
A_H=\frac{2}{3}\pi Q^2
\left(
1+\sqrt{1+\frac{48\tau}{Q^2}}
\right),
\end{eqnarray}
a result which provides a good agreement with 
the numerical data\footnote{The corresponding expressions 
for  $d=5,6$
are much more complicated and not so accurate.
}.


\medskip

\begin{figure}[ht!]
\begin{center}
\includegraphics[height=.26\textheight, angle =0]{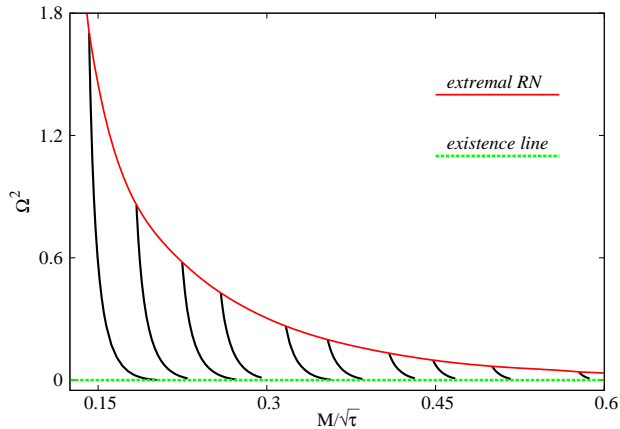} 
\end{center}
  \vspace{-0.5cm}
\caption{ The square of $\Omega^2$
  is plotted as a function of the scaled mass for $d=4$
	solutions, and 
	for several values of the parameter $Q/\sqrt{\tau}$ -- black solid lines -- 
	between 
	$0.1414$ (left)
	and 
	$0.577$ (right).
	A curve with constant $Q/\sqrt{\tau}$ interpolates between  an extremal RN  
	solution and a point on in the existence line, $\Omega=0$. RN BHs with $\Omega^2>0$ are unstable.
}
\label{perturb}
\end{figure}

The RN solutions supporting these {\it zero modes} or {\it  marginally stable mode}
separate different domains
of dynamical 
stability
in the parameter space.
We have investigated this issue for the (physically most interesting) $d=4$ case.
Starting with a more general Ansatz than 
(\ref{metric}), 
(\ref{YMansatz})
with a dependence of both
time and radial coordinate,
which includes more gauge potentials and an extra $g_{rt}$ metric component,
one considers again 
fluctuations around the RN BH, with
a value of  non-Abelian magnetic gauge potential close to the vacuum  value everywhere,
$
 w=\pm 1+ \epsilon w_1(r) e^{\Omega t}+\dots,
$
and real $\Omega$.
Again, it turns out that,
to lowest order in $\epsilon$, the coupled equations separate,
%
$w_1(r)$  being a solution of the equation
\begin{eqnarray}
\label{eq-W}
\frac{d^2 w_1}{d\rho^2 }- \left[\Omega^2+\frac{2 N}{r^2} \left(1-\frac{2\tau Q^2}{r^4}\right) \right]w_1=0,
\end{eqnarray}
where we have introduced a new `tortoise' coordinate  $\rho$ defined by $\frac{d r}{d\rho }=N$,
such that the horizon appears at
$\rho  \to  -\infty$.

 This eigenvalue problem has been solved 
by assuming again that $w_1$ is finite everywhere and vanishes at infinity.
Restricting again to the fundamental mode, 
we display
in Fig. \ref{perturb} the square of the frequency as a function of the mass parameter $M$ 
for several values of $Q$.
One notices that, given $Q$, the RN BH becomes unstable for all values of $M$
below a critical value, or equivalently, when the horizon is sufficiently small. 
Also, the solutions with $\Omega^2\to 0$ corresponds precisely
to the $d=4$ 
{\it existence line} discussed above.

\section{The hairy BH solutions}

The instability of the RN solution found above can be viewed as an indication for 
the existence of a new branch of BH solutions within the EeYM model, having nontrivial magnetic non-Abelian fields outside the
horizon, and continuously connected to the RN solution.\footnote{
This  branching off of a family of solutions at the onset of an instability is a recurrent pattern in BH physics. Examples include the Gregory-Laflamme instability 
of $d\geqslant 5$ 
black strings \cite{Gregory:1993vy,Gubser:2001ac},
the $d=4$ BHs with sinchronyzed hair discussed in the Introduction 
or
 the bumpy BHs in
 $d\geqslant 6$  dimensions
\cite{Dias:2009iu,Dias:2014cia}.
}
This is confirmed by numerical results
obtained for
$4\leqslant d\leqslant 8$,
that we now illustrate.

Let us start our discussion by noticing 
 that the total derivative structure 
of the equation for the electric potential 
(\ref{eq-V})
allows treating the value of the electric charge as an input parameter.
However, 
the
same equation excludes the existence of particle-like
configurations with a regular origin and $Q\neq 0$. 
 Thus, the only physically interesting solutions 
 of this model describe BHs,
 with an event horizon at $r=r_h>0$, located at the largest root of $N(r_h)=0$. 
The metric and the gauge field must be regular at the
horizon, which, in the non-extremal case
 implies an approximate
solution
 around $r=r_h$ of the form
\begin{eqnarray}
\label{exp-eh}
N(r)=N_1(r-r_h)+\dots,
~
\sigma(r)=\sigma_h+\sigma_1(r-r_h)+\dots,
\\
\nonumber
~w(r)=w_h+w_1(r-r_h)+\dots,~
 V(r)=v_1(r-r_h)+\dots,
\end{eqnarray} 
all coefficients being determined by $w_h$ and $\sigma_h$. 
It is also straightforward to show that the requirement of finite energy 
implies the following asymptotic behavior at large $r$ 
\begin{eqnarray}
\label{exp-inf}
 m(r)=M_0-\frac{(d-3)q^2}{2r^{d-3}} +\dots,
~
\sigma(r)=1-\frac{(d-3)^2 J^2}{2r^{2(d-2)}}+\dots,
\\
\nonumber
~ w(r)=\pm 1+\frac{ J}{r^{d-3}}+\dots,~V(r)=\Phi-\frac{q}{r^{d-3}}+\dots~.
\end{eqnarray} 
Once the parameters $\sigma_h,~w_h$ and $J,~M,~q$ are specified, all other coefficients in 
(\ref{exp-eh}), (\ref{exp-inf}) can be computed order by order.
The mass and electric charge of the solutions are given by
$M={\cal V}_{(d-2)}(d-2)M_0/8\pi$, 
$Q={\cal V}_{(d-2)}q/8\pi$.
Other quantities of interest are
the Hawking temperature $T_H= \frac{1}{4 \pi} \sigma(r_h) N'(r_h)$, 
the event horizon area $A_H={\cal V}_{(d-2)}r_h^{d-2}$ and the chemical potential $\Phi$.
The constant $J$ in (\ref{exp-inf}) 
is an order parameter describing the deviation from the Abelian RN solution.

\begin{figure}[ht!]
\begin{center}
\includegraphics[height=.26\textheight, angle =0]{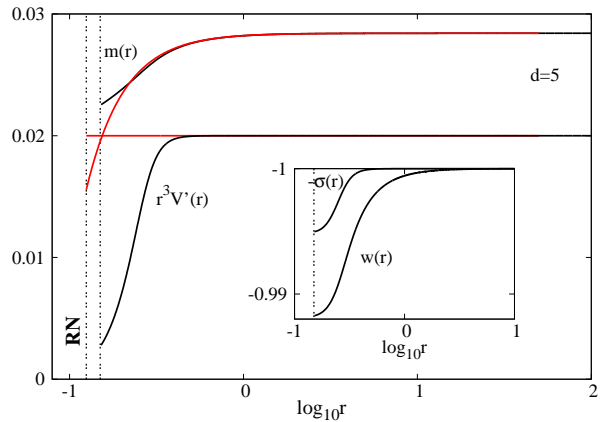} 
\end{center}
  \vspace{-0.5cm}
\caption{ The profiles of a typical $d=5$
non-Abelian solution  (black line) 
is shown
together with the corresponding RN BH (red line) with the same
mass and electric charge parameters,
 $M = 4.186$ and $Q = 2$, respectively.
One notices that the hairy solution is entropically favoured, its 
 horizon radius
being larger than in the RN case.
}
\label{profile}
\end{figure}

 The solutions of the field equations 
interpolating between the  asymptotics (\ref{exp-eh}), (\ref{exp-inf})
were constructed numerically, by employing a shooting strategy.
 For a $F(2)^2+F(4)^2$ model (the only case shown here),
the input parameters are 
 $Q,r_h$ and $\tau$.
Then the solutions are found  for discrete
 values of the parameter $w_h$, labeled by the number of nodes, $n$, of the magnetic
 YM potential $w(r)$.
However, as mentioned above,
  in this work we shall restrict to
  the fundamental set of solutions which possess a monotonic behavior 
of the magnetic gauge potential $w(r)$.

The profile of a typical $d=5$ 
solution is shown in Figure \ref{profile}
(a similar pattern has been found for other spacetime dimensions).
This figure shows that, for the same values of the mass and electric charge,
the RN solution has a smaller event horizon radius (and thus a smaller entropy),
than the non-Abelian BH. Consequently,   as expected, the hairy solutions are entropically favoured.

\begin{figure}[ht!]
\begin{center}
\includegraphics[height=.26\textheight, angle =0]{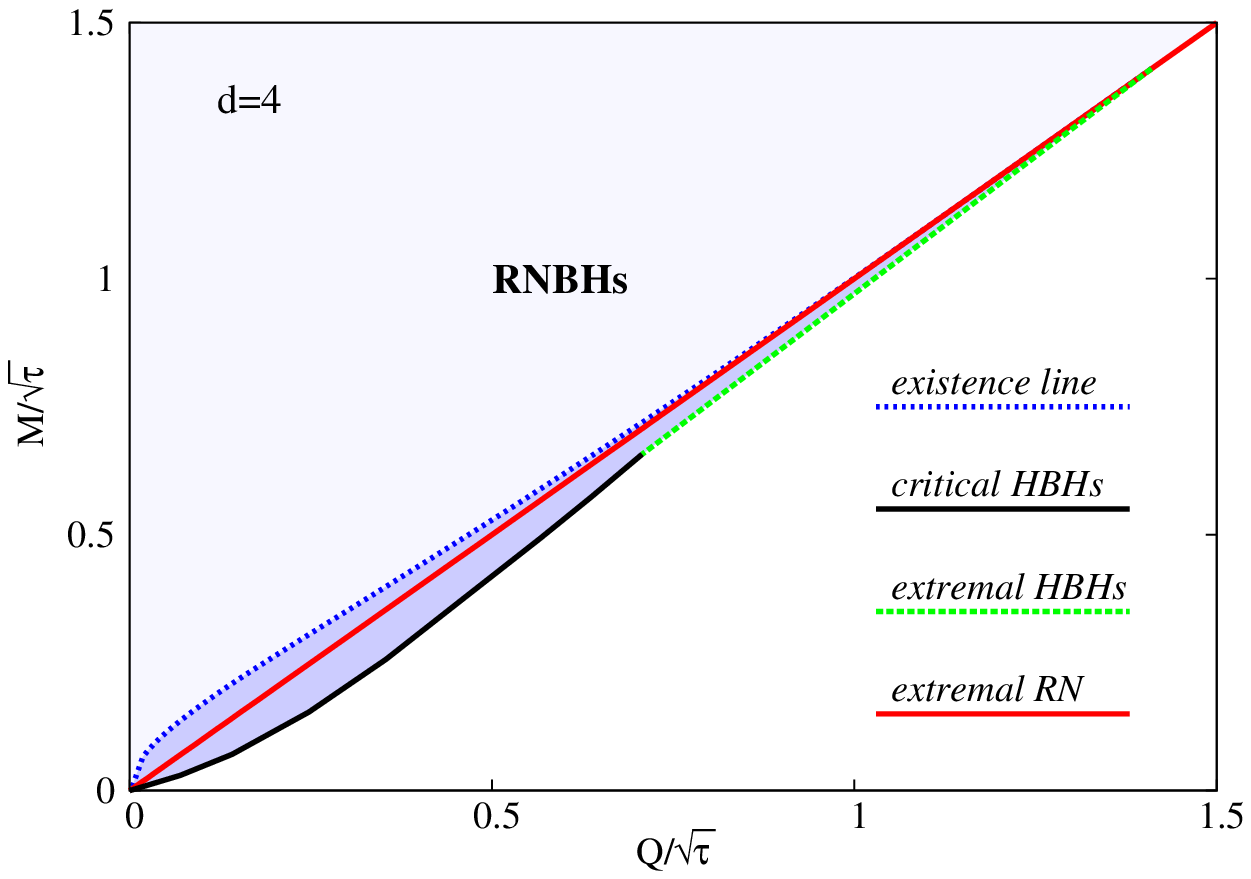} 
\includegraphics[height=.26\textheight, angle =0]{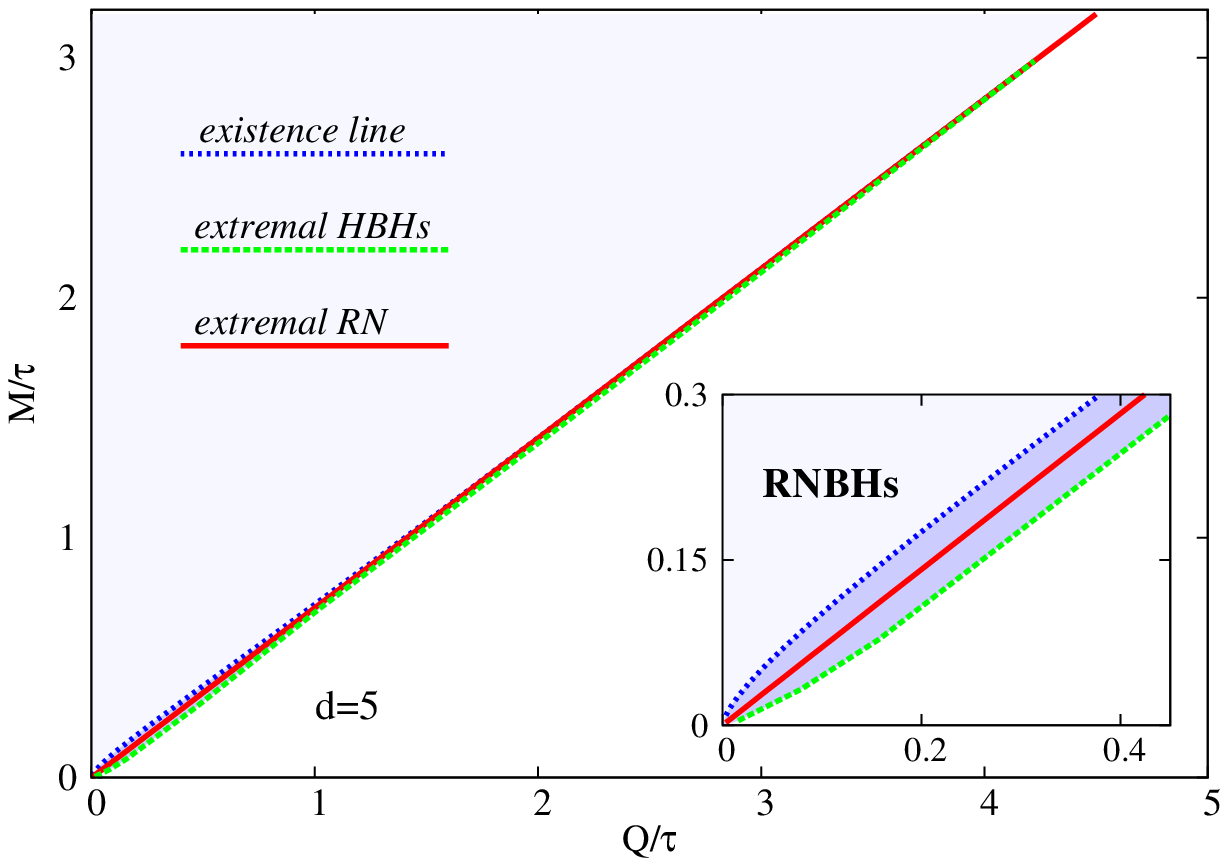} \ \ 
\end{center}
  \vspace{-1.25cm}
\caption{The domain of existence for  $d=4$ and $d=5$ hairy BHs
 (HBHs, shaded dark blue region) in a mass $vs.$ electric charge
 diagram.} 
\label{MQ}
\end{figure}

\begin{figure}[ht!]
\begin{center}
\includegraphics[height=.26\textheight, angle =0]{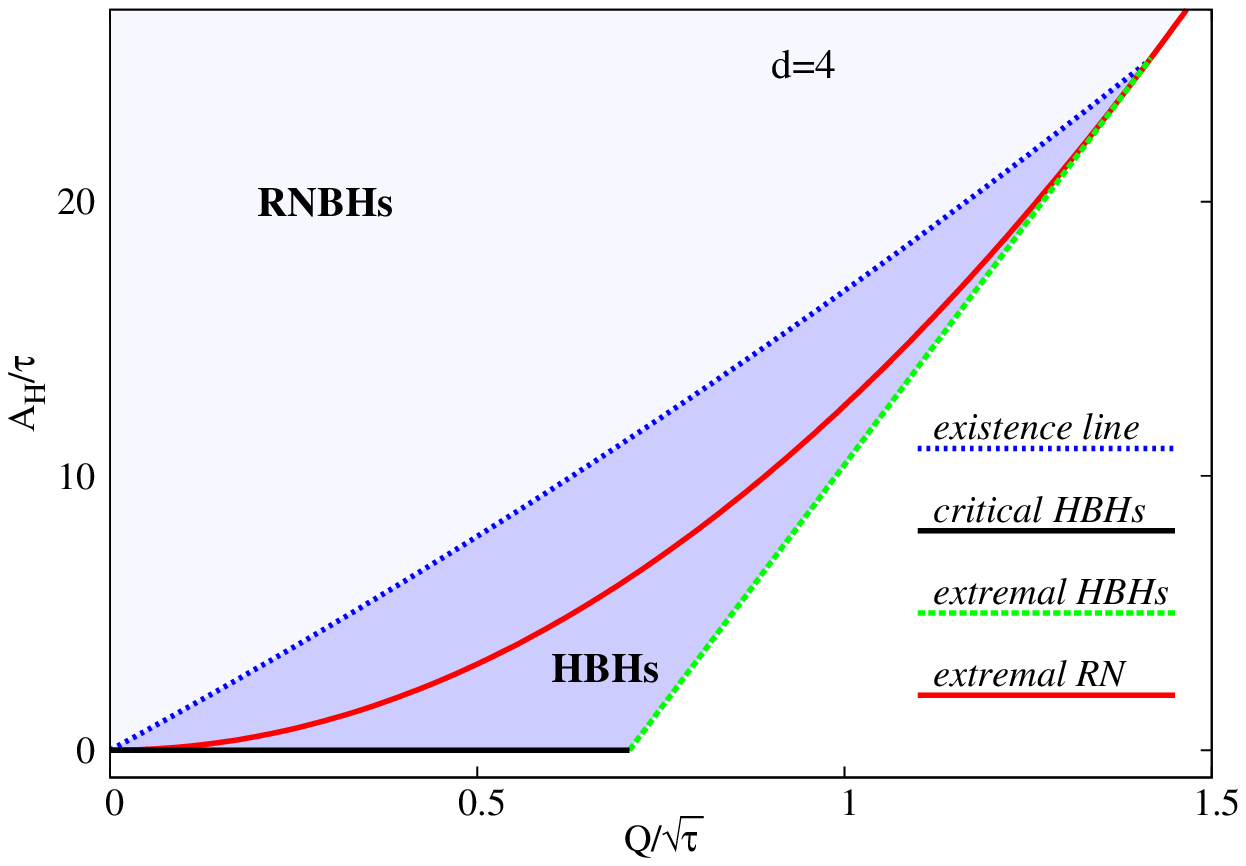} 
\includegraphics[height=.26\textheight, angle =0]{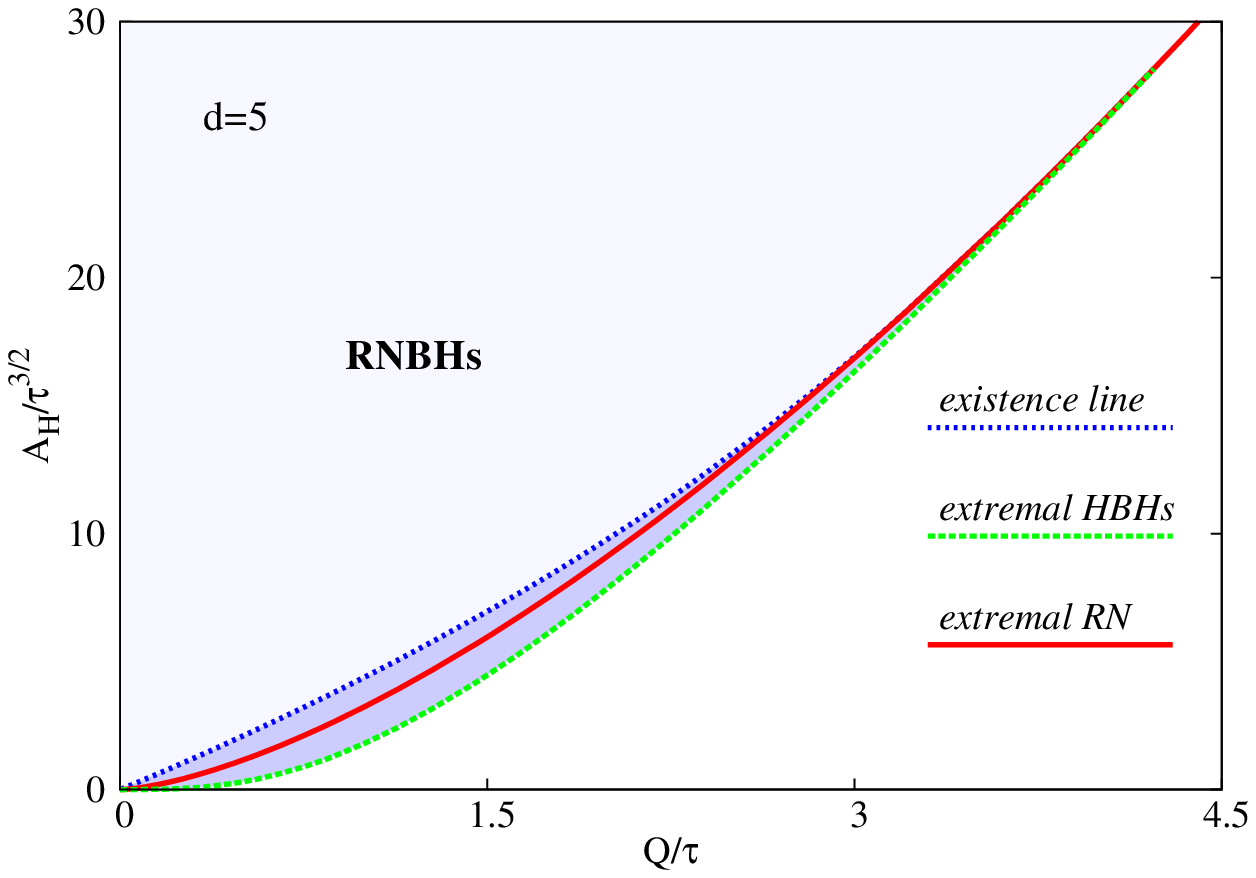} \ \ 
\end{center}
  \vspace{-1.25cm}
\caption{The domain of existence for  $d=4$ and $d=5$ hairy BHs
 (shaded dark blue region) in a event horizon area $vs.$ electric charge
 diagram.} 
\label{AHQ}
\end{figure}

In Figures 
\ref{MQ}, 
\ref{AHQ} 
the mass and  horizon area of all set of $d=4,5$ hairy BH solutions are shown as
functions of the electric charge.\footnote{
The shaded hairy BHs region is obtained by extrapolating to the continuum
the results from a large set of numerical solutions.
The picture for $d=6$ 
(the only other case where we investigated extensively the domain
of existence of hairy BHs)
is very similar to that found for $d=5$.
}
In these plots, the region where the hairy BHs exist is delimited by
{\bf i)} the subset of RN solutions that support the fundamental {\it existence line} of non-Abelian clouds\footnote{ 
This demonstrates that these hairy BHs are the non-linear
realization of non-Abelian clouds.}
(blue dotted line);
{\bf ii)} the set of {\it extremal} ($i.e.$, zero temperature)  hairy BHs (green dashed line).
For $d=4$, the extremal solutions are constructed directly;
for $d=5,6$ they were found by extrapolating
the data for near-extremal configurations.
In four dimensions there is one extra boundary formed by
{\bf iii)} the set of {\it critical} solutions.
These  $d=4$ solutions 
are found by extrapolating the numerical data.
They
possess zero horizon size and appear to be singular,
as found $e.g.$ by evaluating the value of the Ricci scalar at the horizon.

%

This special behaviour can be partially understood
by studying the near horizon limit of the extremal 
hairy BHs.
 The condition of extremality
implies  
$N(r)=N_2 (r-r_h)^2+\dots,$ 
as $r\to r_h$,
while the expansion of 
$w(r)$,
$\sigma(r)$
and $V(r)$ is similar to that in (\ref{exp-eh}).
Then,
restricting for simplicity to a $F(2)^2+F(4)^2$ model,  eqs. (\ref{eq-m})-(\ref{eq-V}) 
reduce to two
 algebraic relations\footnote{As expected, the near horizon structure of the extremal hairy solutions 
can be extended to a full
$AdS_2\times S^{d-2}$ 
exact solution of the field equations,
their properties being essentially fixed by (\ref{exp-extr1}).
}
\begin{eqnarray}
\label{exp-extr1}
&&
(d-2)(d-3)+2(5d^2-37 d+72)Y-4(d-3)X+18(d-4)(d-5)Y^2=0,
\\
&&
\nonumber
\sqrt{\frac{d-2}{d-3}}\frac{1+2Y}{2\sqrt{X}}\sqrt{1+\frac{6Y(d-4)(d-5)}{(d-2)(d-3)}}-\frac{q}{r_h^{d-3}}=0,
~~{\rm with}~~X=\frac{\tau}{r_h^2},~~Y=\tau\frac{(1-w(r_h)^2)^2}{r_h^4}~.
\end{eqnarray}
After eliminating the $w(r_h)$ parameter,  
 one finds\footnote{Although one can write
a general $(Q,A_H)$-relation valid for $d\geqslant 6$,
its expression is very complicated;
however, one finds $Q\to 0$ as $A_H\to 0$.
}
 that the extremal BHs satisfy
 the following 
 charge-area relations:
\begin{eqnarray}
\label{exp-extr2}
&&
\frac{Q}{\sqrt{\tau}}=\frac{1}{\sqrt{2}}
\left(1+\frac{A_H}{8\pi \tau}
\right),~~{\rm for}~~d=4,
\\
&&
\nonumber
\frac{Q}{\tau}=\frac{3}{16 \pi }\sqrt{\frac{3}{2}}
\left(
        \frac{A_H}{\tau^{3/2}}+\frac{4}{3}2^{2/3}\pi^{4/3} \frac{A_H^{1/3}}{\sqrt{\tau}}
\right),~~{\rm for}~~d=5,
\\
&&
\nonumber
\frac{Q}{\tau^{3/2}}=\frac{A_H}{6\sqrt{2} \pi \tau^2}
\left(
\sqrt{ 13+\frac{8\sqrt{6}\pi \tau}{\sqrt{A_H}}}
-2
\right)
\sqrt{1+\sqrt{13+\frac{8\sqrt{6}\pi \tau}{\sqrt{A_H}}}},~~{\rm for}~~d=6.
\end{eqnarray} 
Therefore, the $d=4$ extremal hairy BHs are special, stopping to exist for a minimal value of $Q=\sqrt{ \tau/2}$,
where the horizon area vanishes.
As seen in Figures
\ref{MQ}, 
\ref{AHQ},
the set of {\it critical} solutions connect this point with the 
limiting
configuration with vanishing (scaled) quantities.
No similar solutions are found for $d>4$,
since the limit $Q\to 0$
is allowed in that case.

Let us remark that the 
domains of existence for 
RN BHs and hairy BHs
overlap in a region, see
Figures
\ref{MQ}, 
\ref{AHQ}.
Therein, 
we have found that
 the free energy $F=M-T_HA_H/4$ 
of a hairy solution
is lower than that of the RN configurations with the same 
values for temperature and electric charge.
Finally, we notice the existence of
 overcharged non-Abelian solutions, $i.e$ with electric charge to mass ration greater than unity, 
which do not possess RN counterparts
($e.g.$ for $d=5$
those between the extremal RN  and the extremal hairy BH lines). These solutions cannot arise dynamically from the instability of RN BHs.

\section{Further remarks}
The paradigmatic coloured BHs are disconnected from the RN solution and are unstable against linear perturbations.\footnote{
Strictly speaking, even in the simplest  SU(2)
case, this holds for the static case only.
The spinning solutions necessarily possess an electric charge
\cite{Kleihaus:2000kg,VanderBij:2001nm,Kleihaus:2016rgf}
but their instability has never been established.
}
By considering a simple model with higher order curvature terms of the gauge field (dubbed EeYM model), we have constructed here a qualitatively different set of electrically charged, coloured BHs. The extended YM terms can  provide a
tachyonic mass for the eYM magnetic perturbations 
around the embedded RN BH.
 This leads to the existence of unstable modes. At the threshold of the unstable spectrum lies a zero mode, whose non-linear realization is the family of hairy BHs. The similarity with the more familiar superradiant instability of Kerr BHs is clear, and, as in that case, we expect a dynamical evolution to drive the unstable modes into forming  condensate of non-Abelian magnetic field around the RN BH, and saturating when an appropriate hairy BH forms.

We remark that, for
 $d=5$,
 a rather similar picture is found when 
considering instead solutions in a 
Einstein--Yang-Mills--Chern-Simons model 
\cite{Brihaye:2010wp},
the Chern-Simons term providing an alternative to the higher
order curvature terms of the YM hierarchy employed here.
Again, the hairy BHs emerge as perturbations of the RN solution,
being thermodynamically favoured over the Abelian configurations.

As a possible avenue for future research, it would be interesting to consider
  the stability of the  hairy
solutions in this work.
Since they maximize the entropy for given global charges, we expect them to be stable.
This is indeed confirmed by the $d=4$ results reported in \cite{Radu:2011ip},
which were found, however, for an  SU(3) gauge group. 
The corresponding problem in the  $SO(d+1)$ case
appears to be more challenging and we leave it for future study.

\medskip

Let us close by remarking on some similarities with yet another class of solutions: the coloured, electrically charged BHs in Anti-de Sitter (AdS) spacetime.
As found in \cite{Gubser:2008zu},
the RN-AdS BH becomes unstable when considered as a solution of the pure EYM theory, 
the 'box'-type behaviour of the AdS spacetime provinding the
mechanism for the
appearance of a magnetic non-Abelian cloud close to the horizon. 
Similarly to the situation here,
this feature occurs for a particular set of RN-AdS
configurations which form an {\it existence line} in the parameter space.
Again, 
the hairy BHs are the nonlinear realization of
the non-Abelian clouds.
Their study
via gauge/gravity duality 
 has received a considerable attention
in the literature 
(see $e.g.$ the review 
\cite{Cai:2015cya})
leading  to  
 models of holographic superconductors. 
It would be interesting to  
 explore the possibility that, despite the different asymptotic structure,
 the hairy BHs in this work
could also provide connections to phenomena observed in
condensed matter physics.

\medskip
\medskip

{\bf Acknowledgement}
 C.H. and E. R. acknowledge  funding from the FCT-IF programme.
This work was also partially supported 
by  the  H2020-MSCA-RISE-2015 Grant No.  StronGrHEP-690904, 
and by the CIDMA project UID/MAT/04106/2013.  
\setcounter{equation}{0}


\end{document}